\documentclass[twocolumn,showpacs,10pt,amsmath,amssymb]{revtex4}
\usepackage{graphicx}

\usepackage{amsmath,amsfonts,amssymb,bm}

\begin{document}

\draft
\title{L\'evy walks with variable waiting time: a ballistic case}

\author
{A. Kami\'nska and T. Srokowski}

\affiliation{
 Institute of Nuclear Physics, Polish Academy of Sciences, PL -- 31-342
Krak\'ow, Poland }

\date{\today}

\begin{abstract}
The L\'evy walk process for a lower interval of an excursion times distribution ($\alpha<1$) is discussed. 
The particle rests between the jumps and the waiting time is position-dependent. Two cases are considered: a rising and diminishing 
waiting time rate $\nu(x)$, which require different approximations of the master equation. 
The process comprises two phases of the motion: particles at rest and in flight. The density distributions for them 
are derived, as a solution of corresponding fractional equations. 
For strongly falling $\nu(x)$, the resting particles density assumes the $\alpha$-stable form (truncated at fronts), and the process 
resolves itself to the L\'evy flights. 
The diffusion is enhanced for this case but no longer ballistic, in contrast to the case for the rising $\nu(x)$. 
The analytical results are compared with Monte Carlo trajectory simulations. 
The results
qualitatively agree with observed properties of human and animal movements. 

\end{abstract} 


\maketitle

\section{Introduction}

The ansatz of the L\'evy walk model is a time of flight distribution $\psi(\tau)$ which determines the size of the particle displacement. 
It possesses a slowly decaying asymptotics, $\psi(\tau)\sim\tau^{-1-\alpha}$, 
where $0<\alpha<2$. The case of the lower interval, $\alpha<1$, is usually called a 'ballistic case' since it is characterised by 
a ballistic diffusion: the mean squared displacement rises with time as $t^2$, as a consequence of the infinite mean of $\psi(\tau)$ \cite{zab}. 
Processes characterised by very long tails of the time of flight distribution and ballistic transport are encountered, e.g., for phenomena 
related to nanocrystals \cite{brok,marg}, where the L\'evy statistics, ergodicity breaking and ageing are observed. 
The distribution of a blinking time of quantum dots appears universal indicating a power-law form 
with $\alpha=0.5$ \cite{marg}. The L\'evy walk in the ballistic regime was analysed, e.g., in respect to the ageing phenomena 
and compared with corresponding jumping processes \cite{mag}. 
The L\'evy walk with rests, which includes both the ageing phenomena and the algebraic resting time distribution, was recently 
applied to model a neuronal transport; the experimental data for this process indicate $\alpha=0.52$ \cite{song1}. 
The power-law tails of the distributions are typical for migration problems 
of humans and animals \cite{gei,gon,sca,vis,rhee,raich,reyn,song,boy1}. In particular, the L\'evy index in its lower interval 
was reported in a study of marine predator vertical movements: $\alpha=0.9$ for a leatherback turtle and 
$\alpha=0.7$ for a magellanic penguin \cite{sims}. 

The L\'evy walk model can be generalised by introducing rests at the points of the consecutive velocity renewals \cite{kla2,zab1,tay}. 
The waiting time is given by an independent distribution which may be exponential or possessing long algebraic tails. 
In the latter case, the competition between the L\'evy walk stretches and rests modifies the time dependence of the variance: 
diffusion is no longer ballistic and the two effects may be compensated leading to a normal diffusion \cite{klsok}. 
The heavy tails of the time distribution are observed for the human behaviour (ranging from communication to entertainment and 
work patterns) \cite{bara} and  emerge in an analysis of a population dynamics in the framework of a random networks theory \cite{fedo}. 
Similarly as for the standard L\'evy walk, the walker position is restricted by the total evolution time $t$ 
and a velocity $v$: $|x|<vt$. However, since the walker typically moves in an environment with a differentiated structure, 
some regions may exhibit stronger trapping effects and the waiting time may not have the same distribution in the entire space. 
This is the case for the transport in the disordered systems \cite{bou} and migration of humans and animals \cite{song,boy1}. 
In particular, the human movements estimated from a banknotes flow and discussed in \cite{gei}, depend on environment conditions. Moreover, 
the trapping emerges in the Hamiltonian dynamical systems when a trajectory performs a L\'evy walk in a chaotic environment. It may encounter 
a regular structure in the phase space and then stick to it \cite{lich}; such a structure is selfsimilar. 
The medium heterogeneity can be taken into account in a L\'evy flights model by introducing a variable diffusion coefficient, as 
was done, e.g., for the folded polymers \cite{bro}. The transport in a medium with heterogeneously distributed traps
can be formulated in the framework of a subordination technique where the heterogeneity is taken into account as a position-dependent 
subordinator \cite{sro15}. Then the resulting Fokker-Planck equation possesses a position-dependent diffusion
coefficient and may be of a fractional order both in position and time. The fractional derivative over time reflects long tails 
of the waiting time distribution. In an alternative approach to the problem of the random walk in nonhomogeneous media (with 
long rests), one assumes that the order of the time-derivative (which corresponds to an exponent of the anomalous diffusion) 
is position-dependent \cite{che}. 

The properties of the L\'evy walk model for the waiting time distribution with a position-dependent rate were analysed for the 
subballistic case ($\alpha>1$) \cite{kam}. In the present paper, we extend that analysis to the case $\alpha<1$. In Sec.II, we define the densities 
of two phases of the motion, particles in flight and at rest. They are governed by a master equation which is transformed to fractional 
equations and solved for both the rising and diminishing waiting time rate. Those cases are separately considered in Secs.III and IV, respectively.

\section{General expressions} 

As usual, we define the L\'evy walk as a jumping process for which the (finite) time of a single flight is a random variable and follows from 
a one-sided stable distribution with a power-law asymptotics $\psi(\tau)$; in the following, 
we will consider the case $\alpha<1$. 
The jump density distribution, 
\begin{equation}
\label{jden}
\bar\psi(\xi,\tau)=\frac{1}{2}\delta(|\xi|-v\tau)\psi(\tau), 
\end{equation}
reflects a coupling between the jump size and the time of flight ($v=$ const). 
Before the jump, the particle rests at the point $x$ and the waiting time is random: exponentially distributed 
with a mean given by a function $1/\nu(x)$. 
Therefore, the process consists of two phases, flying and resting, which are described by two density distributions, 
$p_v(x,t)$ and $p_r(x,t)$, respectively. 

The master equation for $p_r(x,t)$ can be obtained from an infinitesimal transition probability $x'\to x$ by the integration 
over all possible $x'$ and times of flight \cite{kam}. It reads, 
\begin{equation}
\label{meq}
\begin{split}
\frac{\partial}{\partial t}&p_r(x,t)= -\nu(x)p_r(x,t)\\ 
&+\int_0^t\int\nu(x') p_r(x',t-t')\frac{1}{2}\psi(t')\delta(|x-x'|-vt')dt'dx'.
\end{split}
\end{equation} 
The density $p_v(x,t)$, in turn, takes into account flights not terminated at time $t$, i.e., particles that are still in flight 
at the point $x$. It is given by the integral, 
\begin{equation}
\label{pv}
p_v(x,t)=\int\int_{0}^t\nu(x')\Psi(t')\delta(|x-x'|-vt')p_r(x',t-t')dx'dt',  
\end{equation} 
where $\Psi(t)$ means a survival probability: $\Psi(t)=\int_t^\infty\psi(t')dt'$. 

The normalisations of the individual density for the both phases, $\phi_r(t)=\int p_r(x,t)dx$ and $\phi_v(t)=\int p_{v}(x,t)dx$, 
are not preserved and depend on time. The time evolution of the normalisation integrals is governed by the following equations, 
\begin{eqnarray}
  \label{q1}
  \frac{\partial}{\partial t}\phi_r(t)&=&-\Phi(t)+\int_0^t\Phi (t-t')\psi(t')dt'\\ \nonumber
  \phi_v(t)&=&\int_0^t\Phi (t-t')\Psi(t')dt',
  \end{eqnarray} 
where $\Phi(t)=\int \nu(x)p_r(x,t)dx$. The limit of small $s$ in the Laplace expansion of the function $\psi(t)$, 
\begin{equation}
\label{psiods}
\psi(s)=1-c_1s^{\alpha}, 
\end{equation} 
differs from the case of the upper interval of $\alpha$, where the leading term is proportional to $s$. 
After taking into account the expansion (\ref{psiods}), the transformed Eq.(\ref{q1}) becomes, 
\begin{eqnarray}
  \label{q2}
  s\phi_r(s)-1&=&-c_1s^\alpha\Phi(s)\\ \nonumber
 \phi_v(s)&=&c_1s^{\alpha-1}\Phi(s). 
  \end{eqnarray} 
Eq.(\ref{q2}) can be easily solved for the case $\nu(x)=$ const=1; after simple algebra and inverting the Laplace transform, 
we obtain the final result in the form of a Mittag-Leffler function \cite{mathai2}, 
\begin{equation}
  \label{mlphi}
 \phi_r(t)=E_{1-\alpha}(-c_{1}t^{1-\alpha}),
\end{equation}
which means that the normalisation integral of the resting phase declines with time as a power-law, $t^{\alpha-1}$, for $t\gg1$. 
$\phi_v(t)$, in turn, rises as $1-\phi_r(t)$. 

We want to solve Eq.(\ref{meq}) and find densities for both phases but the form of the solution and techniques applied 
depend on whether the function $\nu(x)$ is rising or diminishing, which corresponds to declining and 
rising mean waiting time, respectively. We will consider both cases separately. 

\section{Mean waiting time as a diminishing function of position}

In this section, we assume that the walker dwells relatively short in a distant area, i.e., $\nu(x)$ is a rising function. 
The starting point is Eq.(\ref{meq}) which will be analysed for small $s$ and $k$. 
The Fourier transform from that equation reads, 
\begin{equation}
  \label{lmeq}
\begin{split}
\frac{\partial}{\partial t}&p_r(k,t) = -[\nu(x)p_r(x,t)]_F \\
&+ \int_0^t[\nu(x') p_r(x',t-t')]_F \psi(t')\cos(vkt')dt'.
\end{split}
  \end{equation}
Then we take the Laplace transform, apply Eq.(\ref{psiods}) and expand $\cos(vkt')\sim 1-(vkt')^2/2$, which yields, 
\begin{equation}
  \label{lmeq1}
s^{3-\alpha}p_r(k,s)-s^{2-\alpha}P_0(k)=-c_1[s^{2}+Bv^2k^2][\nu(x)p_r(x,t)]_{F-L}, 
  \end{equation}
where $B=\alpha(1-\alpha)/2$ and $P_0(x)$ stands for an initial condition. Since the term $s^{3-\alpha}$ is small compared to 
the term $s^2$, it can be neglected and, for $P_0(x)=\delta(x)$, the Fourier inversion of Eq.(\ref{lmeq1}) yields, 
\begin{equation}
  \label{rfall}
  s^2\nu(x)p_r(x,s)=Bv^2\frac{d^2}{dx^2}[\nu(x)p_r(x,s)]. 
  \end{equation}
Inversion of the Laplace transform produces a wave equation, 
\begin{equation}
  \label{rfal}
  \nu(x)\frac{\partial^2}{\partial t^2}p_r(x,t)=Bv^2\frac{\partial^2}{\partial x^2}[\nu(x)p_r(x,t)]. 
  \end{equation}
The solution of Eq.(\ref{rfall}) is straightforward: 
\begin{equation}
  \label{soll1}
\nu(x)p_r(x,s)=A(s)\exp(-xsb/v),
\end{equation}
where $b=1/\sqrt{B}$ and $A(s)$ is an arbitrary function which is to be determined from the normalisation condition.
Note that neglecting in the expansion in the powers of $k$ the terms higher than $k^2$ changes the front position 
at a given time. This resolves itself to a smaller effective value of the velocity: $v\to v/b$. 
In the following, we set $v=b$. 

The Laplace transform, applied to Eq.(\ref{pv}), yields an equation for the density of particles in flight, 
  \begin{equation}
  \label{solvl0}
\begin{split}
p_v&(k,s)=[c_1s^{\alpha-1}\\
&-c_1s^{\alpha-3}\frac{1}{2}(\alpha-1)(\alpha-2)k^{2}][\nu(x)p_r(x,t)]_{F-L}, 
\end{split}
  \end{equation}
which, after combining with the Fourier transform from Eq.(\ref{rfall}), produces the expression: 
\begin{equation}
  \label{solvl1}
p_v(k,s)=\frac{2c_1}{\alpha}s^{\alpha-1}[\nu(x)p_r(x,t)]_{F-L}.
  \end{equation}
  
To evaluate the function $A(s)$ in Eq.(\ref{soll1}), we utilise the normalisation condition of the total density, 
\begin{equation}
  \label{tnorm}
\phi_r(s)+\phi_v (s)=1/s, 
\end{equation}
where $\phi_r(s)$ is evaluated from the density $p_r(x,t)$ in the form, 
\begin{equation}
\label{prnorm}
p_r(x,t)=\frac{1}{b}\int_0^tA(t-t')\delta(x-t')/\nu(x)dt', 
\end{equation}
according to Eq.(\ref{soll1}). Then Eq.(\ref{tnorm}) becomes, 
\begin{equation}
  \label{tnorm1}
[1/\nu(t)]_L A(s)+\frac{2c_1}{\alpha}s^{\alpha-2}A(s)=1/s.
\end{equation}

To further proceed with the analysis of the above equation, one has to assume a specific form of $\nu(x)$. 
It will be parametrised as a power law, 
\begin{equation}
\label{nuodx}
\nu(x)=|x|^{-\theta}~~~~(\theta>-\alpha), 
\end{equation}
and the parameter $\theta$ serves as a measure of the gradient of the trap density; 
we will demonstrate that predictions of the model (density distributions, relaxation and diffusion properties) 
qualitatively depend on $\theta$. 
The form (\ref{nuodx}) is natural if the environment has a selfsimilar structure \cite{osh,met1} and is often 
observed, e.g., in migration problems. It has been argued \cite{sims} 
that movements of some animals are characterised by the power-law dependences because the prey (e.g., krill) is distributed 
in this way and, for this problem, $\theta>0$. Obviously, the animals abide longer in regions where food is in abundance. 
In this section, we consider the case $\theta<0$. 

For the power-law form of $\nu(x)$, Eq.(\ref{tnorm1}) becomes 
\begin{equation}
  \label{tnorm2}
\Gamma(1+\theta)A(s)+\frac{2c_1}{\alpha}s^{\alpha+\theta-1}A(s)=s^{\theta},
  \end{equation}
which is the Laplace transform of an Abel equation of the second kind. The inversion yields, 
\begin{equation}
  \label{aeq}
\begin{split}
A(t)&+\frac{2c_1}{\alpha\Gamma(1+\theta)\Gamma(1-\theta-\alpha)}\int_0^t\frac{A(t')}{(t-t')^{\alpha+\theta}}dt'\\
&= -\frac{\sin(\pi\theta)}{\pi} t^{-\theta-1},
\end{split}
\end{equation}
and the solution of the above equation is well-known \cite{gormai}; it reads, 
\begin{equation}
  \label{aeq1}
A(t)=-\frac{\sin(\pi\theta)}{\pi} \left[t^{-\theta-1}+\int_0^t(t-t')^{-\theta-1}e'_{1-\alpha-\theta}(t',a)dt'\right],
\end{equation}
where $a=2c_1/\alpha\Gamma(1+\theta)$ and 
\begin{equation}
  \label{aeq2}
\begin{split}
e'_{1-\alpha-\theta}(t',a)&=\frac{d}{dt}E_{1-\alpha-\theta}(-at^{1-\alpha-\theta})\\
&=-t^{-\alpha-\theta}E_{1-\alpha-\theta,1-\alpha-\theta}(-at^{1-\alpha-\theta}).
\end{split}
\end{equation}
Performing the integral in Eq.(\ref{aeq1}) yields the final expression, 
\begin{equation}
  \label{aeq3}
\begin{split}
A(t)&=-\frac{\sin(\pi\theta)}{\pi}[t^{-\theta-1}\\
&-t^{-\alpha-2\theta} E_{1-\alpha-\theta,1-\alpha-2\theta}(-at^{1-\alpha-\theta})].
\end{split}
\end{equation}
For $|\theta|<\alpha$, the second term behaves like $\sim t^{-\theta-1}$, according to the asymptotic form 
of the generalised Mittag-Leffler function $E_{\alpha,\beta}(x)$. This implies a decline of $A(t)$, and, 
consequently, a decay of the resting phase. 

To derive a finite form of the density $p_r(x,t)$, we invert Eq.(\ref{soll1}) and obtain the expression $A(t-x)|x|^\theta$; 
the inserting of the asymptotic form of $A(t)$ yields, 
\begin{equation}
  \label{sol1}
  p_r(x,t)\sim |x|^{\theta}(t-x)^{-\theta-1}.
\end{equation}
Then, if $|x|$ is small compared to $t$, the term $|x|^{\theta}$ determines the position dependence of the density. 
The density of particles in flight follows from Eq.(\ref{solvl1}) which, for the power law $\nu(x)$, becomes, 
\begin{equation}
  \label{solvl2}
p_v(x,s)=\frac{2c_1}{\alpha}\frac{s^{\theta+\alpha-1}}
  {\Gamma(\theta+1)+\frac{2}{\alpha}s^{\theta+\alpha-1}}\exp(-xs).
  \end{equation}
\begin{figure}
\includegraphics[width=90mm]{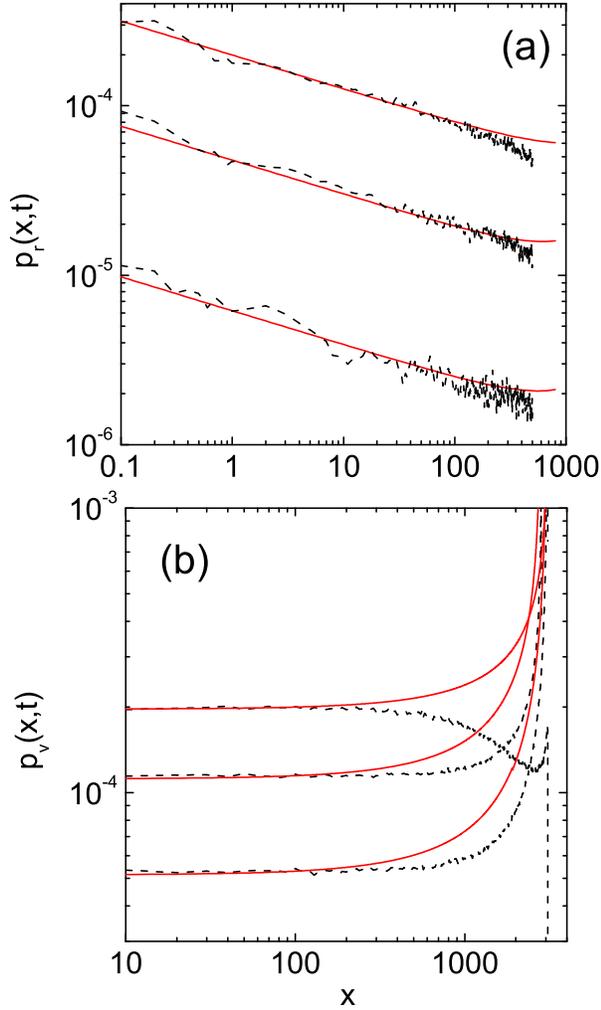}
\caption{Density distributions obtained from Monte Carlo simulations for $t=10^3$ and $\theta=-0.2$. (a) $p_r(x,t)$ for 
$\alpha=0.5$, 0.7 and 0.9 (black dashed lines, from bottom to top). Red solid lines follow from Eq.(\ref{sol1}). 
(b) $p_v(x,t)$ for $\alpha=0.3$, 0.5 and 0.7 (black dashed lines, from bottom to top at left side). 
Red solid lines follow from Eq.(\ref{solv}).}
\end{figure} 
The final expression follows from the inversion of the Laplace transform: 
\begin{equation}
  \label{solv}
\begin{split}
&p_v(x,t)=\frac{2c_1}{\alpha}[\delta(t-x)\\
&+(t-x)^{-\alpha-\theta}E_{1-\alpha-\theta,1-\alpha-\theta}(-a(t-x)^{1-\alpha-\theta})].
\end{split}
  \end{equation}

The intensity of both phases of the motion depends on time and the relaxation of $\phi_r(x)$ can directly be derived from $A(t)$ 
by means of Eq.(\ref{prnorm}) and (\ref{tnorm2}): 
\begin{equation}
\label{phirl}
\phi_r(s)=s^{-\theta-1}A(s)=\frac{1}{s}\frac{s^{1-\alpha-\theta}}{\Gamma(1+\theta)s^{1-\alpha-\theta}+2c_1/\alpha},
\end{equation}
which results in a Mittag-Leffler relaxation pattern, 
\begin{equation}
\label{phir}
\phi_r(t)=\frac{1}{\Gamma(1+\theta)}E_{1-\alpha-\theta}(-\frac{2c_1}{\alpha\Gamma(1+\theta)}t^{1-\alpha-\theta}), 
\end{equation}
with the asymptotics $\phi_r(t)\sim t^{\alpha+\theta-1}$. The intensity of the flight phase, $\phi_v(t)$, rises to unity. 

\begin{figure}
\includegraphics[width=90mm]{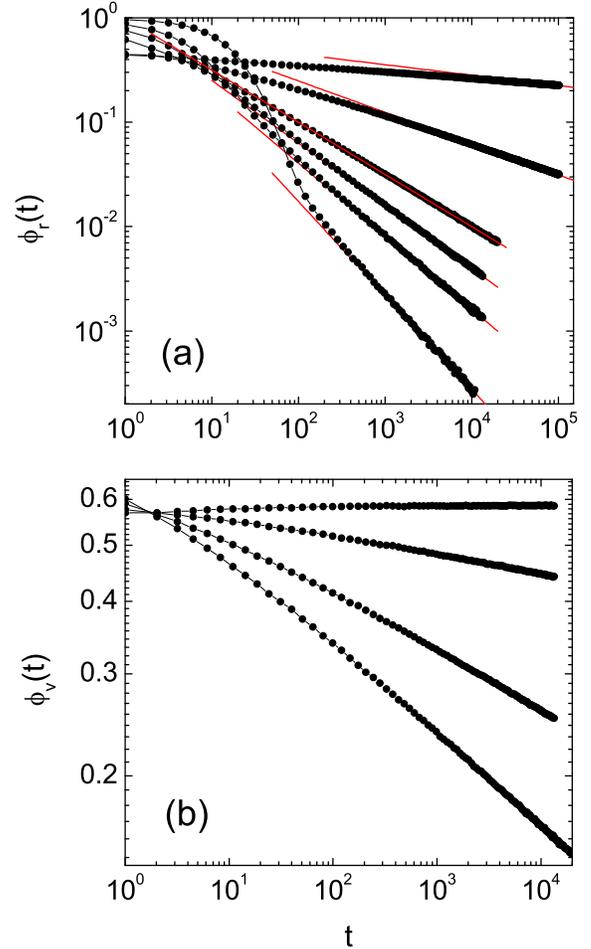}
\caption{(a) Intensity of the resting phase obtained from Monte Carlo simulations for $\alpha=0.5$ and the following 
values of $\theta$: 0.4, 0.2, 0, -0.1, -0.2 and -0.4 (points, from top to bottom at right side). 
Solid red lines mark the dependence $t^{\alpha+\theta-1}$. (b) Intensity of the flying phase for $\alpha=0.5$ 
and the following values of $\theta$: 0.5, 0.6, 0.8 and 1 (points, from top to bottom at right side).}
\end{figure} 

On the other hand, the density distributions were calculated from simulations of individual trajectories by means 
of a Monte Carlo method. The time of flight was sampled from the distribution 
\begin{eqnarray} 
\label{psinu}
\psi(t)=\left\{\begin{array}{ll}
\alpha\epsilon^\alpha t^{-1-\alpha}  &\mbox{for  $t>\epsilon$}\\
0 &\mbox{for  $t\le\epsilon$};
\end{array}
\right.
\end{eqnarray}  
in all calculations, $\epsilon=0.1$ and $v=b$. The waiting time was determined from the exponential 
distribution with the rate (\ref{nuodx}). Fig.1 presents the density distributions of both phases 
of the motion. For a fixed $t$, $p_r(x,t)$ declines according to $x^\theta$ and this dependence persists over a wide range of $x$; the simulation 
results agree with Eq.(\ref{sol1}). 
$p_v(x,t)$, in turn, assumes a constant value if $x$ is not very large. In the latter case, for a large $\alpha$, the density 
falls before the singularity at $x=vt$ emerges. The analytical results, which are compared with the simulations in the figure, 
follow from Eq.(\ref{solv}), where the function $E_{\alpha,\beta}(x)$ was numerically calculated. Both results agree if $x$ is not very large. 

The time-dependence of the moments can be simply determined by a differentiation of the characteristic function of 
the total density $p(x,t)=p_r(x,t)+p_v(x,t)$. For the variance we have \cite{kam}, 
\begin{equation}
\label{e13}
\begin{split}
\langle x^2\rangle(s)&=-\frac{\partial^2}{\partial k^2}p(k,s)|_{k=0}\\
&=2c_{1} (1-\alpha)^2s^{\alpha-3}[\nu(x)p_r(x,t)]_{F-L}(k=0), 
\end{split}
\end{equation} 
which, after the evaluation of the Fourier transform, becomes: 
\begin{equation}
\label{e14}
\langle x^2\rangle(s)=2c_1 (1-\alpha)^2s^{-2}(1/s-\phi_r(s)).  
\end{equation} 
Inserting into the above equation $\phi_r(s)$ from Eq.(\ref{phirl}) and dropping the term that falls faster with $s$, yields 
\begin{equation}
\label{e15}
\langle x^2\rangle(t)\propto t^2. 
\end{equation} 
We conclude that the ballistic diffusion emerges for any $\alpha$ and any $\theta<0$, similarly to a well-known 
result both for both the homogeneous process ($\nu(x)=$ const) and L\'evy walks without rests \cite{zab}.

\section{Mean waiting time as a rising function of position}

In the preceded section, the master equation (\ref{lmeq1}) was handled in the limit of small $s$ for a given $k$: 
the first term (proportional to $s^{3-\alpha}$) was neglected which resulted in the wave equation. However, that procedure 
becomes problematic if $\nu(x)$ is a diminishing function, as the following arguments demonstrate. Let us define 
an auxiliary function $w(x,t)=\nu(x)p_r(x,t)$; then, after inversion of the Fourier transform, Eq.(\ref{lmeq1}) takes the form, 
\begin{equation}
\label{lmeq2}
s^{3-\alpha}\nu(x)^{-1}w(k,s)-s^{2-\alpha}W_0(x)=-c_1(s^{2}+\frac{d^2}{dx^2})w(x,t). 
\end{equation}
It is obvious that if $|x|$ is large and $\nu(x)$ falls sufficiently fast, $s^{1-\alpha}\nu(x)^{-1}$ may not be negligible for large $|x|$. 
First, we will try to assess how fast $\nu(x)$ must fall to prevent neglecting that term. For that purpose, 
we assume $\nu(x)$ in the form (\ref{nuodx}) ($\theta>0$). 
The comparison of typical times characterising the two phases of the motion shows that there is a distinguished threshold value 
of $\theta$, $\theta_{th}=1-\alpha$, which marks a transition from the flying particles dominated process to the predominance 
of the resting phase. The time related to the flying phase can be estimated by a mean time of flight for a given $t$, 
$t_v=\int_0^t t't'^{-1-\alpha}dt'\propto t^{1-\alpha}$, while the time of the resting phase by the waiting time for a given $x$, 
$t_r=|x|^\theta$. If we approximate $|x|$ by the fronts, $|x|=vt$, then $t_r/t_v\propto t^{\alpha+\theta-1}$. 
Therefore, if $0\le\theta<\theta_{th}$ the flying phase prevails at large time, similarly to the case $\theta<0$, and one can expect 
that the results of the section III remain valid. On the other hand, those results (in particular, Eq.(\ref{solv})) 
do not make sense for $\theta>\theta_{th}$. 
The threshold value of $\theta$ is visible in the Monte Carlo simulations presented in Fig.2: $\phi_r(t)$ falls for 
all $\theta<\theta_{th}$ and the pattern agrees with Eq.(\ref{phir}). 

To analyse the case $\theta>\theta_{th}$, we take into account all the terms in Eq.(\ref{lmeq2}). One encounters here 
a well-known problem of the order of limits in $s$ and $k$: the result may depend on this order. In particular, 
the proper asymptotics of the density distributions in the L\'evy walk model (without rests) is achieved when 
the limits $s\to0$ and $k\to0$ are taken simultaneously \cite{schm}, i.e. when $s/k$ converges to a constant, finite value $\kappa$. 
In order to apply this procedure to Eq.(\ref{lmeq1}), we rewrite this equation as, 
\begin{equation}
\label{t6}
  \frac{s^{2-\alpha}}{s^2+k^2}(sp_r(k,s)-P_0(k))=-c_1[|x|^{-\theta}p_r(x,t)]_{F-L}, 
\end{equation}
where the fraction $\frac{s^{2-\alpha}}{s^2+k^2}=|k|^{-\alpha}\frac{\kappa^{2-\alpha}}{\kappa^2+1}\equiv D|k|^{-\alpha}$. Finally, 
\begin{equation}
\label{t7}
sp_r(k,s)-P_0(k)=-D|k|^\alpha[|x|^{-\theta}p_r(x,t)]_{F-L}, 
\end{equation}
and the inversion of the Laplace transform yields, 
\begin{equation}
\label{t8}
\frac{\partial}{\partial t}p_r(k,t)=-D|k|^\alpha[|x|^{-\theta}p_r(x,t)]_F. 
\end{equation}
Note that neglecting the term $s^2$ in Eq.(\ref{lmeq2}) also results in the above equation which observation emphasises the fact that 
the first term is essential for the process. 
Eq.(\ref{t8}) represents the Fourier transformation of a fractional Fokker-Planck equation with the variable diffusion coefficient 
$|x|^{-\theta}$, 
\begin{equation}
\label{t9}
\frac{\partial}{\partial t}p_r(x,t)=D\frac{\partial^\alpha}{\partial |x|^\alpha}[|x|^{-\theta}p_r(x,t)], 
\end{equation}
and describes a Markovian version of a continuous time random walk (CTRW) with the jumping rate $|x|^{-\theta}$ \cite{sro06}. 
In general, the CTRW model \cite{mont} includes a possibility of long rests which results, in the homogeneous case, 
in a fractional derivative over time and the anomalous diffusion \cite{met}. Its coupled form was applied, e.g., in finance \cite{meer}. 
Taking into account only the term $k^0$ in the expansion of the Fourier transform of Eq.(\ref{t9}) yields the asymptotic 
solution ($|k|\ll1$) which agrees with the $\alpha-$stable distribution, 
\begin{equation}
\label{dstab}
p_r(x,t)\sim D\langle|x|^{-\theta}\rangle t|x|^{-1-\alpha}, 
\end{equation}
and corresponds to the L\'evy flights. 
The variable rate of the resting time influences time characteristics of the solution but the distribution shape does not depend on $\theta$. 
\begin{figure}
\includegraphics[width=90mm]{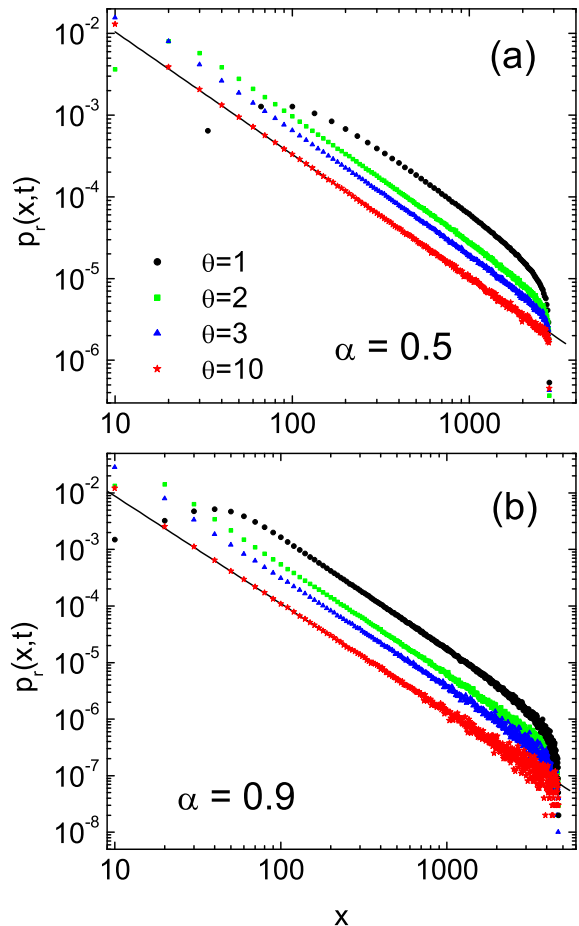}
\caption{$p_r(x,t)$ calculated from trajectory simulations. Solid lines mark the dependence $x^{-1-\alpha}$.}
\end{figure} 
The agreement of Eq.(\ref{dstab}) with CTRW, for which jumps are instantaneous, becomes obvious when one compares typical times of resting 
and flying. For large $\theta$, 
$t_r\gg t_v$ and the time of flight may be neglected. 

The density $p_r(x,t)$, resulting from trajectory simulations, is presented in Fig.3 for two values of $\alpha$. 
It reveals the form $x^{-1-\alpha}$, which comprises the entire range of $x$ if $\theta$ is very large. 
The stable form of the distribution tails abruptly terminates at the fronts and we observe a shape of truncated L\'evy flights 
with a simple cutoff \cite{man}. This form of the distribution is typical, e.g., for migration problems; it represents 
the mobility patterns of humans in such areas as: college campuses, a metropolitan
area, a theme park and a state fair \cite{rhee}. The stable distribution with $\alpha<1$ and an exponential cutoff 
was reported in an analysis of movements of people using data from their mobile phones \cite{gon,sca}. The pattern of human travels 
emerging from the analysis of the bank notes dispersal also reveals the L\'evy statistics in the lower interval of the stability 
index, $\alpha=0.59$ \cite{gei}. The comparison of densities for two values of the parameter $\alpha$, presented in Fig.3, 
demonstrates that the stabile asymptotics comprises the narrower range of $|x|$ if $\alpha$ is smaller; this effect disappears, however, 
for large $\theta$. Then $\nu(x)^{-1}$ strongly rises with the distance, the time of flight may be neglected and the process resembles 
CTRW even for relatively small $|x|$. 

The comparison of Eq.(\ref{sol1}) and (\ref{dstab}) shows that the densities $p_r(x,t)$ for $\theta<0$ and $\theta>\theta_{th}$ are qualitatively 
different (cf. also Fig.1a and Fig.3). This observation does not hold for the density of the flying particles, $p_v(x,t)$, which 
is presented in Fig.4. The plateau, resulting from Eq.(\ref{solv}) and shown in Fig.1b, is still visible but it becomes shorter when $\theta$ rises. 
Instead, $p_v(x,t)$ slowly declines and this behaviour is also visible for 
$\theta<0$ ($\alpha=0.7$ in Fig.1b). Therefore, no clear threshold value of $\theta$ can be recognised in that analysis, 
in contrast to the other observables. 

The time-dependence of $p_r(x,t)$ cannot be determined from the asymptotic solution (\ref{dstab}) since the evaluation 
of the mean in this equation  
requires a knowledge of $p_r(x,t)$ for all $x$. Time-dependence of $\phi_v(t)$ and $\langle x^2(t)\rangle$ was determined from 
the numerical analysis. Fig.2b presents the intensity of the flying phase, $\phi_v(t)$, for some values of $\theta\ge\theta_{th}$. It falls 
as a power-law and converges to a constant for $\theta=\theta_{th}$. The decline of the flying phase is a natural consequence 
of the diminishing of $t_v/t_r$ for $\theta>\theta_{th}$. 

\begin{figure}
\includegraphics[width=90mm]{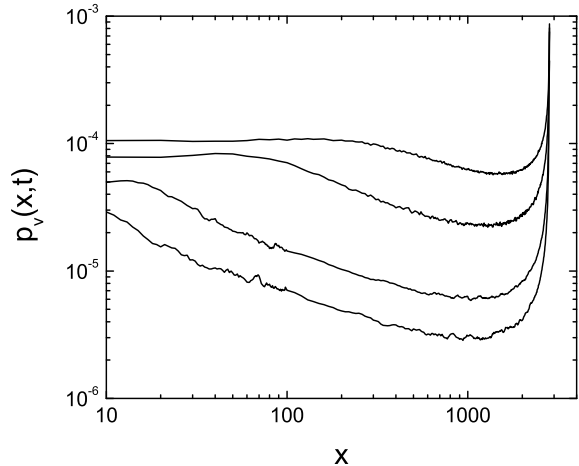}
\caption{$p_v(x,t)$ calculated from trajectory simulations for $\alpha=0.5$ and the following 
values of $\theta$: 0.6, 1, 2 and 3 (from top to bottom).}
\end{figure} 

\begin{figure}
\includegraphics[width=90mm]{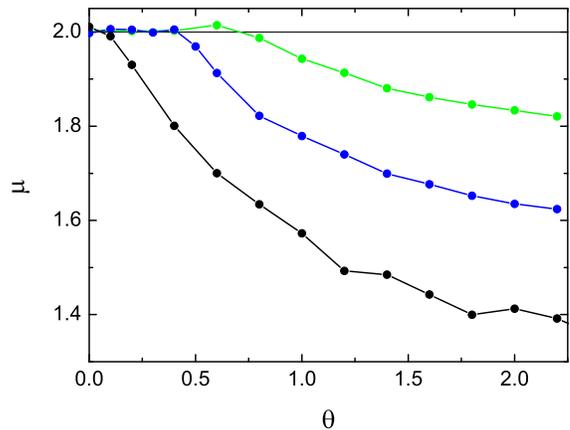}
\caption{The diffusion index $\mu$ (cf. Eq.(\ref{warmu})) as function of $\theta$ for $\alpha=0.3$, 0.6 and 0.9 (from top to bottom).}
\end{figure} 

The analysis of the fluctuations for $\alpha>1$ \cite{kam} shows that the variable waiting time rate essentially influences the diffusion 
pattern. While for the homogeneous case always the enhanced, subballistic behaviour is possible, for the case with the variable rate 
$\nu(x)$, the subdiffusion is also observed. For the present case ($\alpha<1$), 
the Monte Carlo simulations indicate a power-law growth of the variance, 
\begin{equation}
\label{warmu}
\langle x^2\rangle(t)\propto t^\mu, 
\end{equation}
for all $\theta>-\alpha$ and Fig.5 shows the exponent $\mu$ as a function of $\theta$. Since the flying phase prevails for $\theta<\theta_{th}$, 
Eq.(\ref{e15}) is valid there  and the threshold values in the figure agree with $\theta_{th}$. For $\theta>\theta_{th}$, 
all the presented cases are characterised by the enhanced diffusion ($\mu>1$) but the transport is slower than ballistic. 
The decline of the function $\mu(\theta)$ is strongest for large $\alpha$. 
The already mentioned analysis of human movements \cite{rhee} (which indicates the truncated L\'evy statistics) reveals 
such a form of transport: the enhanced diffusion but weaker than ballistic. 
The slowing of the diffusion process due to the resting times is observed in systems homogeneous in space if 
the waiting time has long, power-law tails \cite{klsok}. 

\section{Summary and conclusions}

The analysis of the L\'evy walk process with position-dependent resting times \cite{kam} (defined by the rate $\nu(x)$) 
has been extended to the lower interval of the 
stability index, $\alpha<1$. The shape of the density distributions qualitatively depends on $\nu(x)$. If $\nu(x)$ is a rising 
function, the density of the resting phase, $p_r(x,t)$, is just $1/\nu(x)$ while the density of the flying phase, $p_v(x,t)$, 
assumes a constant value in a wide range 
of $x$. The individual normalisation of both phases is not preserved: the relative intensity of resting particles declines with time. 
On the other hand, the case of diminishing $\nu(x)$ is characterised by $p_r(x,t)$ in the form of the $\alpha$-stable distribution cut off 
at the fronts. This result agrees with the prediction of the Markovian CTRW defined in terms of the stable distribution of the jumping size and 
the waiting time distribution with a variable rate. The stable form of $p_r(x,t)$ is unusual in the L\'evy walk processes but 
comprehensible when one considers typical times of both phases of the motion; since the resting time is very long, 
the time of flight can be neglected and the instantaneous jumping approximates the process well. 

According to the above results, the truncated 
power-law form of the distribution, present, e.g., in patterns of the human movements, is a natural consequence of the heterogeneous environment 
structure which corresponds to the increasing density of traps with the distance: the walker encounters more favoured places. 
Moreover, that form does not emerge for $\alpha>1$ when a stretched-exponential asymptotics always is observed \cite{kam}. 

When $\nu(x)$ rapidly falls, the procedure of neglecting the terms in the master equation which 
are small for $s\to0$, in order to obtain a fractional equation, cannot be applied. 
Decisive for the dynamics, the term appears which its importance owes to the $x$-dependence. 
The proper (compatible with the simulations) form of $p_r(x,t)$ is achieved if the limits $s\to0$ and $k\to0$ are taken simultaneously. 

The diffusion properties of the L\'evy walk process are determined by the integrated density for the flying phase, $\phi_v(t)$, 
which always means the ballistic diffusion if this phase prevails. 
In the opposite case, the numerical calculations indicate the rise of the variance which is slower than ballistic 
but the diffusion remains enhanced. 

The ansatz of the presented random walk model, namely, that the waiting time rate depends on position, 
stems from the observation that the walker moves in an environment that usually possesses 
a structure; this affects the relative time intervals between consecutive displacements. One encounters such complex media when considering, 
e.g., movements of humans and animals. The presented results qualitativelly agree with some features of migration: 
both the density in the form of the truncated L\'evy distribution and the enhanced diffusion weaker than ballistic are observed in the 
migration problems. From the perspective of our formalism, these properties are possible when the resting time distribution depends on position.

\end{document}